\documentclass[sigconf]{acmart}
 
\usepackage[most]{tcolorbox}
\usepackage[ruled,vlined]{algorithm2e}
\usepackage{algpseudocode}
\usepackage{amsmath}
\usepackage{enumitem}
\usepackage{bbm}
\usepackage{pifont}
\usepackage{tabularx}

\newcommand{\cmark}{\ding{51}} 
\newcommand{\xmark}{\ding{55}} 

\newcommand{\best}[1]{\textbf{#1}}
\newcommand{\secondbest}[1]{\underline{#1}}

\newcommand{\green}[1]{\textcolor{green!70!black}{#1}}
\newcommand{\red}[1]{\textcolor{red}{#1}}

\AtBeginDocument{%
  }

\copyrightyear{2026}
\acmYear{2026}
\setcopyright{cc}
\setcctype{by}
\acmConference[ICTIR '26]{Proceedings of the 2026 International ACM SIGIR Conference on Innovative Concepts and Theories in Information Retrieval (ICTIR)}{July 25, 2026}{Melbourne, VIC, Australia}
\acmBooktitle{Proceedings of the 2026 International ACM SIGIR Conference on Innovative Concepts and Theories in Information Retrieval (ICTIR) (ICTIR '26), July 25, 2026, Melbourne, VIC, Australia}
\acmDOI{10.1145/3805713.3820414}
\acmISBN{979-8-4007-2600-2/2026/07}

\begin{document}

\title{Rank, Don’t Generate:\\Statement-level Ranking for Explainable Recommendation}

\author{Ben Kabongo}
\orcid{0009-0005-3411-4262}
\affiliation{%
  \institution{Sorbonne Université}
  \department{CNRS, ISIR}
  \city{Paris}
  \country{France}
}
\email{ben.kabongo@sorbonne-universite.fr}

\author{Arthur Satouf}
\orcid{0009-0001-0198-3868}
\affiliation{%
  \institution{Mila - Quebec AI Institute} 
  \institution{Air Liquide} 
  \institution{Université Paris-Saclay} 
  \city{Montreal}
  \country{Canada}
}
\email{arthur.satouf@gmail.com}

\author{Vincent Guigue}
\orcid{0000-0002-1450-5566}
\affiliation{%
  \institution{AgroParisTech}
  \department{UMR MIA Paris-Saclay}
  \city{Palaiseau}
  \country{France}
}
\email{vincent.guigue@agroparistech.fr}

\renewcommand{\shortauthors}{Ben Kabongo, Arthur Satouf, Vincent Guigue}

\begin{abstract}
Textual explanations, generated with large language models (LLMs), are increasingly used to justify recommendations. Yet, evaluating these explanations remains a critical challenge. Lexical $n$-gram metrics struggle with paraphrases; semantic metrics reward similarity without ensuring factual grounding; LLM-based evaluators are highly prompt-dependent, often proprietary, and hard to replicate, while current evaluation protocols provide little insight into which explanatory factors are faithful and relevant. Meanwhile, proposed models tend to hallucinate unsupported claims or default to generic rationales.

We advocate a shift in objective: \emph{rank, don't generate}. We formalize explainable recommendation as a statement-level ranking problem, where systems rank candidate explanatory statements derived from reviews and return the top-$k$ as explanation. 
This formulation mitigates hallucination by construction and enables fine-grained factual analysis. It also models factor importance through relevance scores and supports standardized, reproducible evaluation with established ranking metrics.
Meaningful assessment, however, requires each statement to be \textit{explanatory} (item facts affecting user experience), \textit{atomic} (one opinion about one aspect), and \textit{unique} (paraphrases consolidated), which is challenging to obtain from noisy reviews.

We address this with (i) an LLM-based extraction pipeline producing explanatory and atomic statements, and (ii) a scalable, semantic clustering method consolidating paraphrases to enforce uniqueness.
Building on this pipeline, we introduce \textsc{StaR}, a benchmark for statement ranking in explainable recommendation, constructed from four Amazon Reviews 2014 product categories. We evaluate popularity-based baselines and state-of-the-art models under \textit{global-level} (all statements) and \textit{item-level} (target item statements) ranking. Popularity baselines are competitive in global-level ranking but outperform state-of-the-art models on average under item-level ranking, exposing critical limitations in personalized explanation ranking. Code, data, and supplementary materials are released to support reproducible research and future work: \url{https://github.com/BenKabongo25/Statement_Ranking}.
\end{abstract}

\ccsdesc[500]{Information systems~Learning to rank}
\ccsdesc[500]{Information systems~Top-k retrieval in databases}
\ccsdesc[300]{Information systems~Language models}
\ccsdesc[300]{Information systems~Clustering}
\ccsdesc[500]{Information systems~Recommender systems}
\ccsdesc[100]{Information systems~Sentiment analysis}
\ccsdesc[300]{Information systems~Clustering and classification}

\keywords{Explainable Recommendation,
Statement-level Ranking,
Evaluation and Benchmarking,
Large Language Models (LLMs),
Statement Extraction,
Semantic Clustering}

\begin{teaserfigure}
\centering
  \vspace{-4mm}
  \includegraphics[width=0.75\linewidth]{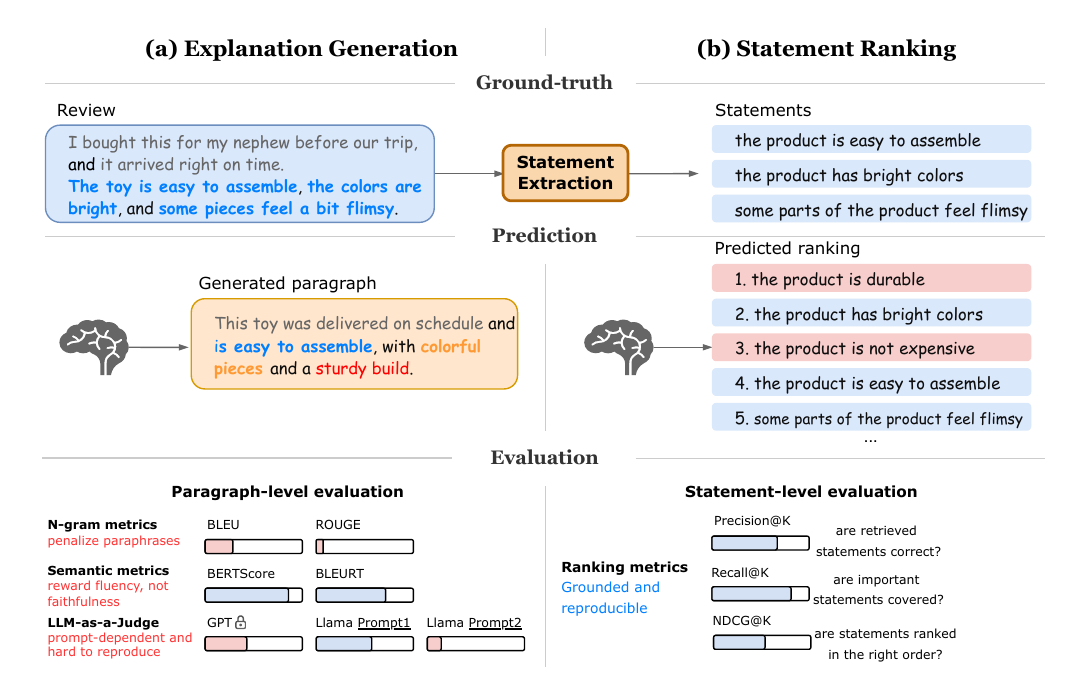}
  \vspace{-4mm}
  \caption{Better explanation evaluation needs better units: rank statements, not paragraphs.}
  \Description{Better explanation evaluation needs better units: rank statements, not paragraphs.}
  \label{fig:teaser}
\end{teaserfigure}

\maketitle

\section{Introduction}
Recommender systems play a central role in helping users navigate large catalogs of content, products, or services by delivering personalized suggestions.
However, the growing reliance on deep learning architectures and techniques~\cite{geng2022recommendation, he2020lightgcn, rajput2023recommender} has made these systems increasingly opaque and difficult to explain or interpret.
Explainable recommendation addresses this issue by providing human-understandable justifications for why an item is recommended, which is crucial for building user trust and improving system transparency.
Among existing paradigms, textual explanations are particularly appealing.
Recent methods~\cite{kabongo2025elixir, li2021personalized, li2023personalized, li2025g, ma2024xrec, xie2023factual} increasingly leverage large language models (LLMs)~\cite{dubey2024llama, team2024gemma, yang2025qwen3} to generate natural-language rationales.
However, reliably assessing the quality of generated explanations remains a critical challenge.

Early evaluation protocols relied on lexical $n$-gram overlap metrics~\cite{banerjee2005meteor, lin2004rouge, papineni2002bleu}, which struggle with synonymy and paraphrasing.
Learned semantic metrics~\cite{sellam2020bleurt, yuan2021bartscore, zhang2019bertscore} reduce dependence on surface form, but semantic similarity does not guarantee factual grounding in the underlying evidence~\cite{honovich2022true, kabongo2025factual}.
LLM-based evaluators~\cite{fu2023gptscore, liu2023g} can provide richer judgments, but are highly prompt-dependent and often proprietary, making them hard to reproduce~\cite{sheng2025analyzing, xu2025investigating}.
Beyond evaluation, generation itself introduces additional failure modes: explainable models may hallucinate unsupported claims or default to generic rationales rather than interaction-specific justifications~\cite{kabongo2025factual, li2021extra, xie2023factual}.
Moreover, explanations are intrinsically multi-faceted: multiple factors may jointly justify a recommendation and vary in importance for a given user (e.g., cast versus storyline for movies), whereas current protocols typically score whole paragraphs holistically, offering limited insight into which factors are supported and failing to reflect their relative importance.

\textbf{Statement-level ranking for explainable recommendation.}
These limitations motivate reframing text-based explainable recommendation as a \emph{ranking} problem over explanatory passages, which we call \textit{statements}.
Following prior work~\cite{li2021extra, li2023relationship, wei2023expgcn}, we advocate a paradigm shift: \textit{rank, don’t generate}. Instead of producing a free-form paragraph, the system ranks candidate statements for each user--item interaction, from the most to the least relevant, and returns the top-$k$ as the explanation.
This formulation offers three key advantages.
First, it decomposes explanations into verifiable units grounded in user reviews, mitigating hallucination and enabling fine-grained faithfulness analysis.
Second, ranking captures relative factor importance via relevance scores, allowing the system to prioritize the most important aspects for a given interaction.
Third, it enables standardized, reproducible evaluation with established ranking metrics (e.g., Precision, Recall, and NDCG~\cite{jarvelin2002cumulated}), supporting objective comparisons across methods.

Despite its appeal, statement-level ranking faces fundamental challenges in dataset construction and evaluation.
In particular, meaningful assessment requires each statement to satisfy three properties:
(i) \textit{explanatoriness}, describing general item facts that may affect user experience rather than circumstantial or personal information with no explanatory value;
(ii) \textit{atomicity}, expressing a single opinion about a single item aspect; and
(iii) \textit{uniqueness}, ensuring that (multiple) semantically equivalent declarations are grouped into a single target to avoid skewing metrics.
Existing datasets typically consist of free-form reviews that interleave explanatory content with noise, merge multiple aspects, and vary widely in style, making it difficult to extract statement sets that meet these requirements.

To address these challenges, prior work introduces \textsc{Extra}~\cite{li2021extra}, which extracts frequent review phrases as statements and deduplicates paraphrases using lexical $n$-gram similarity.
In practice, purely lexical clustering often fails to consolidate semantically equivalent paraphrases expressed with different wording, and extraction from raw reviews remains challenged by non-explanatory and non-atomic content.
Satisfying the above statement properties therefore requires robust extraction to filter noise while producing explanatory, atomic units, as well as semantics-aware clustering to group paraphrases beyond surface-level similarity.

\textbf{Our work.}
We propose improved methods for statement extraction and paraphrase clustering.
For extraction, we introduce a two-stage LLM pipeline: (i) \textit{candidate extraction}, which produces explanatory, atomic statements while filtering review noise, and (ii) \textit{verification}, which removes non-compliant outputs and semantic duplicates.
For clustering, we propose a scalable, semantics-driven procedure: (i) \textit{approximate nearest-neighbor search} retrieves semantically similar candidates using dense embeddings, (ii) \textit{pairwise filtering} removes false positives with a cross-encoder, and (iii) \textit{refinement} builds a similarity graph from validated pairs, extracts connected components as initial clusters, and further splits low-cohesion components to enforce intra-cluster consistency.

Building on this pipeline, we introduce \textsc{StaR} (\textbf{Sta}tement \textbf{R}anking), a new benchmark for statement ranking in explainable recommendation, constructed from four categories (\textit{Toys}, \textit{Clothing}, \textit{Beauty}, and \textit{Sports}) of the Amazon Reviews 2014 dataset~\cite{ni2019justifying}. This yields augmented datasets tailored to statement ranking, with 115K--294K interactions and 718K--1.3M user--item--statement triplets after clustering.
We validate \textsc{StaR} by measuring the required statement properties: human evaluation on sampled examples and LLM-based assessment confirm explanatoriness and atomicity, while unsupervised clustering metrics confirm uniqueness through effective paraphrase consolidation.
Finally, we provide qualitative examples highlighting the limitations of \textsc{Extra} for statement extraction and quantitatively assess the weaknesses of its lexical clustering, demonstrating the advantages of our extraction and clustering procedures.

We introduce simple popularity-based baselines and evaluate them on \textsc{StaR} alongside state-of-the-art explanation ranking models (\textsc{BPER+}~\cite{li2023relationship} and \textsc{ExpGCN}~\cite{wei2023expgcn}), and analyze the impact of key hyperparameters across datasets.
Unlike prior work, we adopt two complementary evaluation paradigms: \textit{global-level} ranking, where candidates are drawn from the full statement universe, and \textit{item-level} ranking, where candidates are restricted to statements associated with the target item.
We find that \textsc{ExpGCN} achieves the strongest performance under global-level ranking, while simple item- and global-frequency signals remain surprisingly competitive.
In contrast, under item-level ranking, a user-history popularity baseline surpasses ($+0.08$ in NDCG@10 on \textit{Toys}) or matches \textsc{ExpGCN}, revealing a clear personalization gap.
Overall, these results expose persistent limitations in personalized explanation ranking and underscore the lack of correct, reproducible evaluation protocols.

In summary, our main contributions are:
\begin{itemize}[leftmargin=*]
    \item We propose an improved LLM-based statement extraction method that produces \emph{explanatory} and \emph{atomic} statements from reviews.
    \item We propose a scalable and semantic statement clustering method that consolidates paraphrases, enforcing \emph{uniqueness}.
    \item We introduce \textsc{StaR}, a new benchmark for statement-level ranking in explainable recommendation, constructed by applying our pipeline to four categories of the Amazon Reviews 2014 dataset.
    \item We introduce popularity-based baselines and benchmark them against state-of-the-art models under both global-level and item-level ranking, and we study the impact of key hyperparameters on ranking performance.
\end{itemize}

\section{Related Work}

\subsection{Text-based Explainable Recommendation}
Explainable recommendation aims to justify why an item is recommended, improving system transparency and user trust~\cite{zhang2020explainable}.
Textual explanations have gained increasing attention, as they present justifications in natural language.
Early approaches relied on \emph{templates} filled with features and opinion words~\cite{li2020generate, tao2019fact, wang2018explainable, zhang2014explicit}, but are inflexible and lack diversity.
Subsequent work directly generates the user's \emph{review} or \emph{tip} as explanation, using progressively advanced architectures: recurrent neural networks (Att2Seq~\cite{dong2017learning}, NRT~\cite{li2017neural}), 
then non-pretrained Transformers (PETER~\cite{li2021personalized}, CER~\cite{raczynski2023problem}), and finally pretrained language models (PEPLER~\cite{li2023personalized}, PRAG~\cite{xie2023factual}).
However, reviews often contain noise and non-explanatory passages that should not be reproduced.
Recent methods (XRec~\cite{ma2024xrec}, G-refer~\cite{li2025g}) therefore transform reviews into explicitly explanatory paragraphs, leveraging LLMs~\cite{achiam2023gpt, dubey2024llama, team2024gemma, yang2025qwen3} to produce increasingly fluent explanations.
Nevertheless, despite improvements in \emph{fluency}, reliably evaluating the actual explanatory quality of generated texts remains critical and challenging.

\subsection{Evaluation of Textual Explanations}
Automatically measuring the quality of generated text remains difficult, including for textual explanations in recommendation.

Early protocols relied on lexical $n$-gram metrics such as BLEU~\cite{papineni2002bleu} and ROUGE~\cite{lin2004rouge}, which struggle with synonymy and paraphrasing and correlate weakly with human judgments.
Widely adopted in review-generation settings, these metrics primarily reward stylistic overlap rather than explanatory usefulness.
More recent work uses semantic metrics, either embedding-based~\cite{zhang2019bertscore} or learned~\cite{sellam2020bleurt, yuan2021bartscore}, which are more robust to paraphrases.
However, semantic similarity does not guarantee factual consistency with the underlying evidence~\cite{honovich2022true, kabongo2025factual}.
LLM-based evaluators~\cite{fu2023gptscore, liu2023g} can better align with human judgments, but reliance on proprietary models raises accessibility concerns, and they are highly prompt-dependent and susceptible to biases, making results hard to reproduce~\cite{sheng2025analyzing, xu2025investigating}.

Moreover, explanation models may degenerate into generic rationales~\cite{li2021extra} or hallucinate unsupported claims~\cite{kabongo2025factual}, highlighting weak factual grounding.
More broadly, paragraph-level evaluation offers limited diagnostic value and neither identifies which explanatory factors are supported by the evidence nor reflects their relative importance.

\subsection{Explanation Ranking Benchmarks and Methods for Explainable Recommendation}
The limitations and lack of uniformity in text-based recommendation evaluation motivated the \textsc{Extra} benchmark~\cite{li2021extra}, which standardizes evaluation using established ranking metrics. 
\textsc{Extra} constructs candidate passages from user reviews, identifying two challenges: reviews contain noise requiring filtering of non-explanatory content, and paraphrases must be consolidated into unique statements to support collaborative filtering and meaningful evaluation.
For extraction, \textsc{Extra} retains recurrent sentences while filtering those with personal pronouns, but this heuristic can retain irrelevant content and discard informative explanations.
For clustering, they use Locality-Sensitive Hashing (LSH)~\cite{anand2011mining} with $n$-grams, which fails to capture semantic similarity beyond lexical overlap, severely limiting quality.

Various models have been proposed for ranking explanations: \textsc{BPER+}~\cite{li2023relationship} decomposes the ternary user--item--statement relation via two matrix factorizations and enriches representations with BERT~\cite{devlin2019bert}; \textsc{ExpGCN}~\cite{wei2023expgcn} divides the heterogeneous graph into homogeneous subgraphs and learns task-oriented representations via graph convolution; and \textsc{TriEOR}~\cite{zhang2023triple} models the ternary relation by decomposing rating prediction as aggregated preference scores.
However, these models are evaluated exclusively in a global-level setting with all statements as candidates.

We introduce popularity-based baselines and extend evaluation to item-level ranking, where candidates are restricted to target item statements, demonstrating that popularity signals are strong for personalized explanation ranking.


\subsection{LLM-based Information Extraction and Retrieval}
LLMs have shown strong capabilities in natural language understanding~\cite{achiam2023gpt, bai2023qwen, dubey2024llama, yang2025qwen3} and are increasingly used for information extraction tasks such as summarization~\cite{zhang2024benchmarking}, named entity recognition~\cite{wang2025gpt}, and  sentiment analysis~\cite{wang2023chatgpt}.
\cite{kabongo2025factual} proposes an LLM-based approach to extract statement--topic--sentiment triplets from reviews to measure factuality in text-based explainable recommendation.
LLMs are also central to modern information retrieval, enabling dense retrieval with semantic embeddings~\cite{zhang2025qwen3, zhao2024dense} and reranking with cross-encoders~\cite{sun2023chatgpt, zhang2025qwen3}.
These advances motivate our use of LLMs for statement extraction and clustering.

\section{Statement-level Explanation Ranking}
Statement ranking for explainable recommendation consists in ordering candidate explanatory sentences—called \textit{statements}—from those that best explain a user--item interaction to those that explain it the least, returning the top-$k$ as the explanation.
We first discuss the statement properties required for meaningful and reproducible evaluation, then formalize the ranking task under two settings: global-level and item-level, and describe the evaluation protocol.

\subsection{Statement Properties}
\label{sec:properties}
To ensure meaningful evaluation with ranking metrics, statements must satisfy three properties: \textit{explanatoriness}, \textit{atomicity}, and \textit{uniqueness}.

\subsubsection*{\textbf{Explanatoriness}}
Each statement should express general, item-related facts or opinions that affect user experience (e.g., \emph{the product is easy to use}), excluding circumstantial details or personal information with limited explanatory value (e.g., \emph{the product arrived while I was on vacation}).

\subsubsection*{\textbf{Atomicity}}
Each statement should express a single opinion about a single item aspect.
A non-atomic statement like \emph{the product is fun and easy to use} cannot be safely matched to two atomic statements (\emph{the product is fun}, \emph{the product is easy to use}) in evaluation, nor can it be correct if only one component is supported.

\subsubsection*{\textbf{Uniqueness}}
Each candidate statement should convey a distinct idea, with paraphrases consolidated into a canonical representation.
Without this, near-duplicates in the top-$k$ (e.g., \emph{the product is fun} and \emph{the product is enjoyable}) distort evaluation. 

\subsection{Statement Ranking Problem Formulation}
Following prior work~\cite{li2021extra, li2023relationship, wei2023expgcn, zhang2023triple}, we formalize statement ranking for explainable recommendation, considering two settings: \emph{global-level} ranking, where candidates are drawn from the full dataset, and \emph{item-level} ranking, where candidates are restricted to the recommended item.


\subsubsection{Definitions and Notation}
Let $\mathcal{U}$ and $\mathcal{I}$ denote the sets of users and items, respectively.
An interaction between a user $u \in \mathcal{U}$ and an item $i \in \mathcal{I}$ is associated with a set of explanatory statements
$\mathcal{S}_{ui}=\{s_1,\ldots,s_{n_{ui}}\}$ that justify the interaction. Ratings may be available, but are not used in the remainder of this paper.
Given $(u,i)$ as input, $\mathcal{S}_{ui}$ constitutes the ground-truth explanation set.
Each statement $s$ may additionally be annotated with a polarity label $p_s \in \{\textsc{pos},\textsc{neg},\textsc{neu}\}$, indicating whether it conveys a positive, negative, or neutral opinion in the context of the interaction.
We define the statements associated with an item as
$\mathcal{S}_i = \bigcup_{u \in \mathcal{U}} \mathcal{S}_{ui}$,
and the global statement universe as
$\mathcal{S} = \bigcup_{u \in \mathcal{U},\, i \in \mathcal{I}} \mathcal{S}_{ui}$.
While statement-level annotations may exist in curated datasets, they are rare in practice.
Accordingly, we introduce in the next sections a procedure to automatically extract statements from user reviews.


Given a candidate set for interaction $(u,i)$,
$\mathcal{S}^{\mathrm{cand}}_{ui} \subseteq \mathcal{S}$,
statement ranking aims to assign a relevance score $\hat{r}_{u,i,s}$ (predicted by a given method) to each candidate statement
$s \in \mathcal{S}^{\mathrm{cand}}_{ui}$ and return the top-$k$ statements as the explanation:
\begin{align}
    \pi_{ui} &:= \mathrm{argsort}_{s \in \mathcal{S}^{\mathrm{cand}}_{ui}} \, \hat{r}_{u,i,s}, \notag \\
    \mathrm{Top}_k(u,i,\mathcal{S}^{\mathrm{cand}}_{ui})
    &:= \{\pi_{ui}(1), \ldots, \pi_{ui}(k)\}.
\end{align}
Here, $\pi_{ui}$ is the permutation of candidate statements sorted by decreasing predicted relevance $\hat{r}_{u,i,s}$, and $\pi_{ui}(j)$ denotes the statement at rank $j$.
Therefore, $\mathrm{Top}_k(u,i,\mathcal{S}^{\mathrm{cand}}_{ui})$ is the set of the $k$ highest-scoring statements returned to explain interaction $(u,i)$.
We assume $\mathcal{S}_{ui} \subseteq \mathcal{S}^{\mathrm{cand}}_{ui}$, i.e., the ground-truth statements are contained in the candidate set.

\subsubsection{Ranking Levels}
The definition of the candidate set $\mathcal{S}^{\mathrm{cand}}_{ui}$ determines the ranking level. We consider two settings: \emph{global-level} and \emph{item-level} ranking.

\subsubsection*{\textbf{Global-level ranking}}
In global-level ranking, we set $\mathcal{S}^{\mathrm{cand}}_{ui}=\mathcal{S}$ and rank all statements in the dataset for each interaction.
This setting enables models to exploit the full statement space and capture broad relationships between users, items, and explanations.
A limitation, however, is that it can produce \emph{item-agnostic} explanations. 

\subsubsection*{\textbf{Item-level ranking}}
In item-level ranking, we set $\mathcal{S}^{\mathrm{cand}}_{ui}=\mathcal{S}_i$ and rank only the statements associated with item $i$.
By construction, this yields \emph{item-specific} explanations and isolates the challenge of ordering candidates that are directly relevant to the item.

\subsection{Statement Ranking Evaluation}
Consider an interaction $(u, i)$, with ground-truth explanatory statements $\mathcal{S}_{ui}$ and candidate set $\mathcal{S}^{\mathrm{cand}}_{ui}$.
Let $\mathrm{Top}_k(u,i,\mathcal{S}^{\mathrm{cand}}_{ui})$ denote the list of the top-$k$ ranked statements, and let $\pi_{ui}(j)$ be the statement at rank $j$.
We evaluate ranking quality using standard information retrieval metrics: Precision (P@k), Recall (R@k), and Normalized Discounted Cumulative Gain (NDCG@k)~\cite{jarvelin2002cumulated}.
We define binary relevance at rank $j$ as:
\begin{align*}
    \mathrm{rel}_j = \delta\!\left(\pi_{ui}(j) \in \mathcal{S}_{ui}\right),
\end{align*}
where $\delta(\cdot)$ is the indicator function and $\mathrm{rel}_j$ indicates whether the statement returned at position $j$ belongs to the ground-truth set $\mathcal{S}_{ui}$.
The metrics are then:
\begin{align}
    \mathrm{P}@k(u,i) &= \frac{1}{k} \sum_{j=1}^{k} \mathrm{rel}_j,
    \quad
    \mathrm{R}@k(u,i) = \frac{1}{|\mathcal{S}_{ui}|} \sum_{j=1}^{k} \mathrm{rel}_j, \notag \\
    \mathrm{NDCG}@k(u,i) &= \frac{1}{Z_k} \sum_{j=1}^{k} \frac{2^{\mathrm{rel}_j}-1}{\log_2(j+1)},
    \quad
    Z_k = \sum_{j=1}^{k} \frac{1}{\log_2(j+1)}
\end{align}

\section{\textsc{StaR} Benchmark}
This section introduces \textsc{StaR}, a benchmark for statement-level ranking in explainable recommendation.
We describe our two-stage LLM-based extraction pipeline that produces \emph{explanatory} and \emph{atomic} statements from reviews, our scalable and semantic-oriented clustering procedure that enforces \emph{uniqueness} by consolidating paraphrases, and the resulting benchmark constructed from four Amazon Reviews 2014 categories.

\begin{figure}[h]
    \vspace{-4mm}
    \centering
    \includegraphics[width=0.8\linewidth]{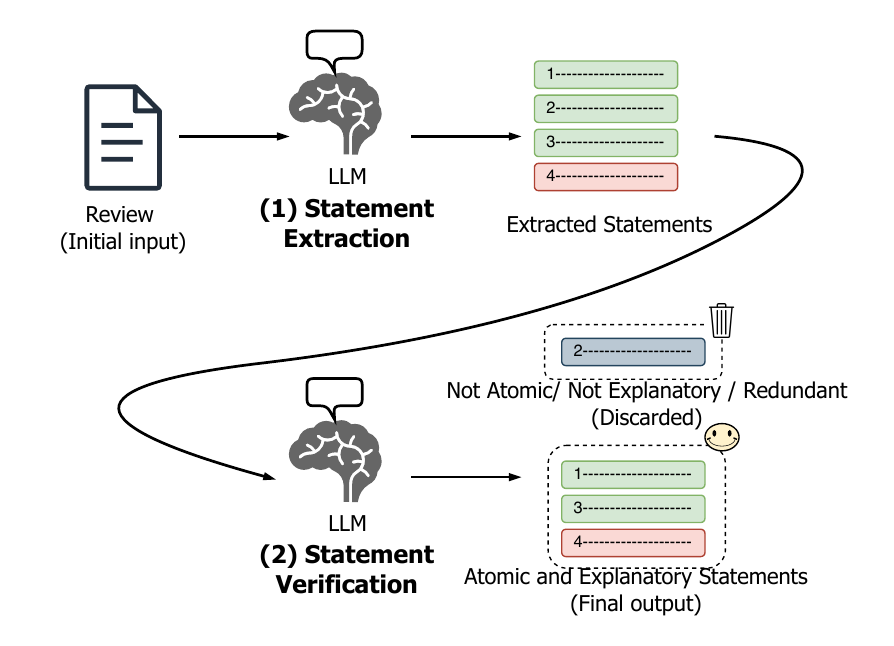}
    \vspace{-5mm}
    \caption{Statement extraction and verification pipeline.
(1) An LLM extracts candidate statements from a raw review.
(2) A second LLM filters non-explanatory, non-atomic, or redundant statements, retaining only compliant outputs.}
    \vspace{-5mm}
    \label{fig:statement_extraction}
    \Description{Statement extraction and verification pipeline.
(1) An LLM extracts candidate statements from a raw review.
(2) A second LLM filters non-explanatory, non-atomic, or redundant statements, retaining only compliant outputs.}
\end{figure}
\begin{figure*}[ht]
    \centering
    \includegraphics[width=\linewidth]{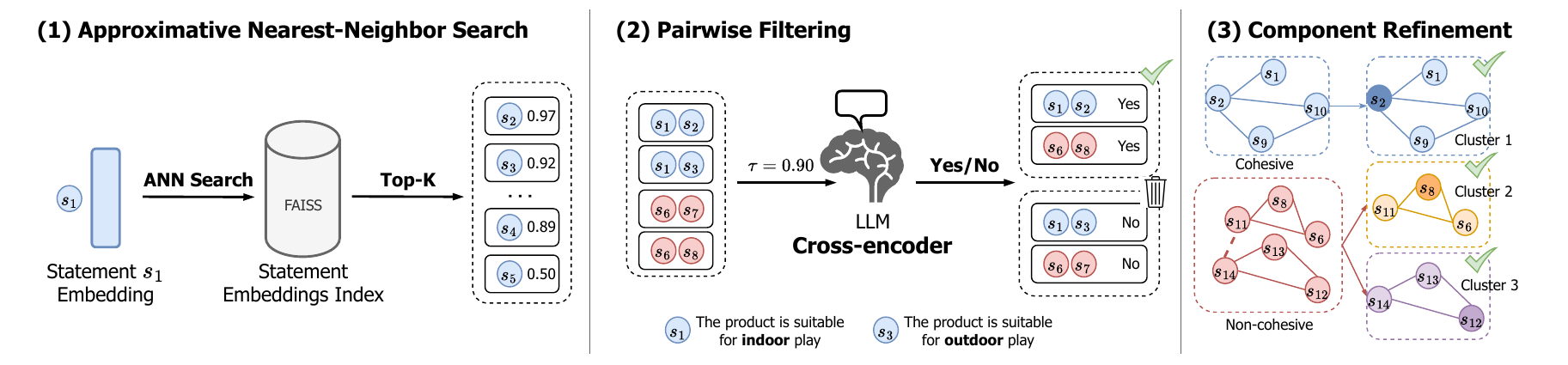}
    \vspace{-6mm}
    \caption{Statement clustering pipeline.
(1) ANN retrieves the top-$K$ semantically similar candidates per statement using dense embeddings.
(2) A cross-encoder re-evaluates candidate pairs and retains only high-confidence matches.
(3) Validated pairs define a similarity graph; connected components form initial clusters, which are refined by splitting low-cohesion components.
}
    \label{fig:statement_clustering}
    \Description{Statement clustering pipeline.
(1) ANN retrieves the top-$K$ semantically similar candidates per statement using dense embeddings.
(2) A cross-encoder re-evaluates candidate pairs and retains only high-confidence matches.
(3) Validated pairs define a similarity graph; connected components form initial clusters, which are refined by splitting low-cohesion components.
}
    \vspace{-2mm}
\end{figure*}
\subsection{Statement Extraction and Verification}
User reviews are a rich source of explanatory evidence for user--item interactions, but they often include noise that does not directly justify the interaction.
To support meaningful evaluation, statement extraction must produce candidates that satisfy two key properties: \emph{explanatoriness} and \emph{atomicity}, while \emph{uniqueness} is enforced in the subsequent clustering stage.

We therefore adopt a two-step pipeline for extracting statements from reviews: (i) \textit{candidate extraction}, where an LLM extracts candidate statements and assigns sentiment labels, and (ii) \textit{verification}, where a second LLM filters candidates to retain only explanatory and atomic statements, as illustrated in Figure~\ref{fig:statement_extraction}.

\subsubsection{Statement Candidate Extraction}
The goal of this step is to produce candidate statements that are \emph{explanatory} and \emph{atomic}, while discarding content that does not directly justify the user--item interaction.
Because sentiment is central to explainable recommendation, we explicitly preserve polarity by extracting statements together with their sentiment labels.
Formally, given a review text $t_{ui}$ written by user $u$ for item $i$, we extract a set of candidate statements
$\hat{\mathcal{S}}_{ui}=\{\hat{s}_1,\ldots,\hat{s}_{m_{ui}}\}$.

\subsubsection{Statement Verification}
Despite careful prompting, a small fraction of extracted candidates remain non-explanatory, non-atomic, or redundant.
We therefore apply a verification stage to improve statement quality by filtering out such outputs.
Formally, given the candidate set $\hat{\mathcal{S}}_{ui}$, verification produces the final statement set
$\mathcal{S}_{ui}=\{s_1,\ldots,s_{n_{ui}}\}$.

\subsubsection*{\textbf{Implementation}}
We use \texttt{Qwen3-14B}~\cite{yang2025qwen3}%
~\footnote{\url{https://huggingface.co/Qwen/Qwen3-14B}} for statement candidate extraction, and a smaller model, \texttt{Qwen3-8B}~\cite{yang2025qwen3}%
~\footnote{\url{https://huggingface.co/Qwen/Qwen3-8B}}, for verification.
Across both stages, we experimented with multiple prompts and model choices, and retained the best-performing configuration.
The final prompts are provided in the supplementary materials.

\subsection{Statement Clustering}
Explanatoriness and atomicity are enforced during statement extraction from reviews.
To complete the evaluation protocol, paraphrased statements must be consolidated to satisfy the third requirement, \emph{uniqueness}.
Given the potentially large number of statements, exhaustively testing semantic equivalence for all pairs is computationally infeasible, and purely lexical matching often misses paraphrases expressed with different wording.

To cluster paraphrases at scale, we use the three-step semantic pipeline shown in Figure~\ref{fig:statement_clustering}:
(i) \textit{approximate nearest-neighbor search} retrieves the top-$K$ semantically similar candidates for each statement,
(ii) \textit{pairwise filtering} forms and re-evaluates candidate pairs with a cross-encoder,
(iii) \textit{refinement} forms a similarity graph from validated pairs, clusters via connected components, and splits low-cohesion clusters.

In the following, we denote by $\mathcal{S}^0$ the set of statements produced by the extraction stage, prior to clustering.

\subsubsection{Approximate Nearest-Neighbor Search}
The goal of this step is to retrieve, for each statement, a small set of semantically closest candidate statements.
Given a statement $s \in \mathcal{S}^0$ and a text encoder $\Phi$, we compute a dense representation $\mathbf{s}=\Phi(s)\in\mathbb{R}^d$, where $d$ is the embedding dimension.
We then perform approximate nearest-neighbor (ANN) search in the embedding space to efficiently identify semantically related candidates.
For each statement $s$, we restrict the search to statements with the same polarity label $p_s$ (to avoid matching opposite sentiments), retrieve the top-$K$ nearest neighbors by cosine similarity, and denote by $\mathrm{ANN}(s;K)$ the resulting set of top-$K$ candidate statements most similar to $s$.

\subsubsection*{\textbf{Implementation}}
To obtain semantically meaningful representations, we use \texttt{Qwen3-Embedding-0.6B}~\cite{zhang2025qwen3}~\footnote{\url{https://huggingface.co/Qwen/Qwen3-Embedding-0.6B}}
as encoder. We apply $\ell_2$ normalization to all statement embeddings, forcing the embedding norm to 1. For each dataset, we embed all statements and index them with Faiss~\cite{douze2025faiss}, building one index per polarity. We set $K = 128$. 

\subsubsection{Pairwise Filtering}
Embedding proximity alone does not guarantee paraphrase equivalence.
We therefore filter retrieved neighbors in two stages.
First, we form candidate pairs from retrieved neighbors and retain only highly similar pairs with cosine similarity above $\tau_{\mathrm{pair}}$:
\begin{align*}
\mathcal{E}^0
=
\bigcup_{s \in \mathcal{S}^0}
\left\{
\{s,s'\}:\;
s' \in \mathrm{ANN}(s;K),\;
\text{\textsc{cos}}(\mathbf{s},\mathbf{s}') \ge \tau_{\mathrm{pair}}
\right\}.
\end{align*}
Second, we re-evaluate each pair with a binary cross-encoder $\Psi$ and keep only pairs predicted as paraphrases, yielding the validated set $\mathcal{E}$.
These validated pairs are then used to build statement clusters in the next step.

\subsubsection*{\textbf{Implementation}}
We set $\tau_{\mathrm{pair}}=0.9$ to retain only highly similar pairs while keeping the number of candidates manageable for the more expensive re-evaluation step.
We use the LLM-based binary cross-encoder \texttt{Qwen3-Reranker-0.6B}~\cite{zhang2025qwen3}~\footnote{\url{https://huggingface.co/Qwen/Qwen3-Reranker-0.6B}} and classify a pair as a paraphrase if its predicted probability exceeds $0.9$, prioritizing high-precision matches.
We experimented with multiple prompts and retained the best-performing one, which is provided in the supplementary materials.

\subsubsection{Refinement}
From the statements $\mathcal{S}^0$ and validated paraphrase pairs $\mathcal{E}$, we build an undirected similarity graph $G=(\mathcal{S}^0,\mathcal{E})$.
Connected components provide initial paraphrase clusters, but may still contain noise because many intra-component pairs were not explicitly validated.
We therefore apply a lightweight refinement step that avoids exhaustive pairwise re-scoring.
For each component $G_\ell=(\mathcal{S}_\ell,\mathcal{E}_\ell)$, we measure cohesion by the minimum intra-component cosine similarity and declare it cohesive only if it exceeds $\tau_{\mathrm{intra}}$; cohesive components are kept as final clusters.
Otherwise, we refine the component with Algorithm~\ref{alg:refine_component}, which partitions it around high-degree \emph{pivots} and merges consecutive pivot clusters when pivot similarity exceeds $\tau_{\mathrm{remerge}}$.
For each resulting cluster, we select a representative statement as the most central element in embedding space (maximizing average similarity to other cluster members), and define the canonical statement set $\mathcal{S}$ as the collection of these representatives.

In our experiments, we set $\tau_{\mathrm{intra}}=0.85$ as the minimum intra-cluster similarity threshold and $\tau_{\mathrm{remerge}}=0.90$ for pivot merging, to maintain high intra-cluster cohesion.

\begin{algorithm}[t]
\small
\SetAlgoSkip{smallskip}
\caption{Refinement of Non-Cohesive Component}
\label{alg:refine_component}
\KwIn{$G_\ell=(\mathcal{S}_\ell,\mathcal{E}_\ell)$, embeddings $\{\mathbf{s}\}_{s\in\mathcal{S}_\ell}$, threshold $\tau_{\mathrm{remerge}}$}
\KwOut{Refined clusters $\mathcal{C}_\ell$}
\textbf{Notations:} $\mathcal{N}_\ell(s)=\{s' \in \mathcal{S}_\ell : (s,s') \in \mathcal{E}_\ell\}$\;
$\mathcal{C}_\ell \leftarrow \emptyset$, $\mathcal{R} \leftarrow \mathcal{S}_\ell$, $\mathcal{A} \leftarrow \emptyset$, $\mathcal{P} \leftarrow \emptyset$\;
\While{$\mathcal{R} \neq \emptyset$}{
  $p \leftarrow \arg\max_{s\in \mathcal{R}} | \mathcal{N}_\ell(s)\cap \mathcal{R}|$,
  $\mathcal{B} \leftarrow \{p\}\cup(\mathcal{N}_\ell(p)\cap \mathcal{R})$\;
  \uIf{$\mathcal{A}=\emptyset$}{
    $\mathcal{A} \leftarrow \mathcal{B}$, $\mathcal{P} \leftarrow \{p\}$\;
  }
  \uElseIf{$\min_{p'\in \mathcal{P}}\text{\textsc{cos}}(\mathbf{p}, \mathbf{p}') \ge \tau_{\mathrm{remerge}}$}{
    $\mathcal{A} \leftarrow \mathcal{A}\cup \mathcal{B}$, $\mathcal{P} \leftarrow \mathcal{P}\cup\{p\}$\;
  }
  \Else{
    $\mathcal{C}_\ell \leftarrow \mathcal{C}_\ell \cup \{\mathcal{A}\}$, $\mathcal{A} \leftarrow \mathcal{B}$, $\mathcal{P} \leftarrow \{p\}$\;
  }
  $\mathcal{R} \leftarrow \mathcal{R} \setminus \mathcal{B}$\;
}
$\mathcal{C}_\ell \leftarrow \mathcal{C}_\ell \cup \{\mathcal{A}\}$\;
\Return $\mathcal{C}_\ell$\;
\end{algorithm}
\begin{table}[t]
\centering
\caption{Dataset statistics for \textsc{StaR} benchmark. We report the reduction rate in \% after clustering in brackets.}
\label{tab:dataset_stats}
\vspace{-4mm}
\resizebox{1\linewidth}{!}{
\begin{tabular}{@{}lrrrr@{}}
\toprule
& \textbf{Toys} & \textbf{Clothes} & \textbf{Beauty} & \textbf{Sports} \\
\midrule
\textbf{Basic Statistics} \\
\;\; Users & 18 594 & 39 371 & 22 361 & 35 594 \\
\;\; Items & 10 130 & 22 940 & 12 086 & 18 322 \\
\;\; Interactions & 155 992 & 274 635 & 198 225 & 294 513 \\
\;\;\;\;\;\;\;\; \textit{Train samples} & 119 052 & 196 458 & 153 587 & 223 528 \\
\;\;\;\;\;\;\;\; \textit{Eval samples} & 18 525 & 39 201 & 22 330 & 35 534 \\
\;\;\;\;\;\;\;\; \textit{Test samples} & 18 415 & 38 976 & 22 308 & 35 451 \\
\midrule
\textbf{Pre-Clustering} \\
\;\; Unique statements & 472 843 & 570 376 & 507 691 & 942 442 \\
\;\;\;\;\;\;\;\; \textit{Positive} & 159 466 & 216 542 & 222 455 & 370 554 \\
\;\;\;\;\;\;\;\; \textit{Negative} & 98 229 & 146 886 & 112 446 & 193 553 \\
\;\;\;\;\;\;\;\; \textit{Neutral} & 215 148 & 206 948 & 172 790 & 378 335 \\
\;\; Min/Avg/Max per inter. & 1/4.60/26 & 1/4.18/26 & 1/4.33/26 & 1/4.65/26 \\
\;\; Min/Avg/Max per item & 3/63.19/1101 & 5/46.35/1099 & 8/67.41/1649 & 8/71.08/3032 \\
\;\; Min/Avg/Max per user & 1/34.24/1701 & 1/27.79/536 & 2/37.45/1115 & 1/37.80/1758 \\
\;\; Total triplets & 718 313 & 1 149 483 & 858 733 & 1 371 531 \\
\midrule
\textbf{Post-Clustering} \\
\;\; Unique statements & 281 664 \green{(40.4\%)} & 260 477 \green{(54.3\%)} & 229 448 \green{(54.8\%)} & 556 209 \green{(41.0\%)} \\
\;\;\;\;\;\;\;\; \textit{Positive}  & 80 476 \green{(49.5\%)} & 84 071 \green{(61.2\%)} & 81 774 \green{(63.2\%)} & 187 055 \green{(49.5\%)} \\
\;\;\;\;\;\;\;\; \textit{Negative}  & 65 637 \green{(33.2\%)} & 75 635 \green{(48.5\%)} & 59 144 \green{(47.4\%)} & 134 118 \green{(30.7\%)} \\
\;\;\;\;\;\;\;\; \textit{Neutral}  & 135 551 \green{(37.0\%)} & 100 771 \green{(51.3\%)} & 88 530 \green{(48.8\%)} & 235 036 \green{(37.9\%)} \\
\;\; Min/Avg/Max per inter. & 1/4.57/26 & 1/4.15/26 & 1/4.28/26 & 1/4.62/25 \\
\;\; Min/Avg/Max per item & 3/56.81/877 & 5/42.49/770 & 7/58.38/1149 & 7/64.67/2081 \\
\;\; Min/Avg/Max per user & 1/32.51/1364 & 1/26.57/470 & 2/35.57/909 & 1/36.74/1526 \\
\;\; Total triplets & 712 963 & 1 142 256 & 850 332 & 1 361 389 \\
\bottomrule
\end{tabular}
}
\vspace{-4mm}
\end{table}
\begin{table*}[htbp]

\caption{Statement ranking results on \textsc{StaR}. We report P/R/NDCG at $k\in\{5,10\}$ for global-level and item-level settings. \textit{Improv.} (for \textsc{BPER+} and \textsc{ExpGCN}) is the relative gain over the best popularity baseline per metric. Significance levels (paired $t$-test against the best baseline): $^{*}$~$p < 0.05$, $^{**}$~$p < 0.01$, $^{***}$~$p < 0.001$.
}
\vspace{-2mm}
\begin{center}
\resizebox{\linewidth}{!}{
\begin{tabular}{lcccccccccccc}
\toprule
& \multicolumn{6}{c}{\textbf{Global-level}} & \multicolumn{6}{c}{\textbf{Item-level}} \\
\cmidrule(lr){2-7} \cmidrule(lr){8-13}
\textbf{Methods} & 
\text{P@5} & \text{R@5} & \text{N@5} & 
\text{P@10} & \text{R@10} & \text{N@10} & 
\text{P@5} & \text{R@5} & \text{N@5} & 
\text{P@10} & \text{R@10} & \text{N@10}
\\
\midrule
\multicolumn{13}{c}{\textbf{Toys}} \\
\midrule
  \textsc{Random} & {0.00238} & {0.00284} & {0.00469} & {0.00119} & {0.00285} & {0.00439} & {0.07441} & {0.09108} & {0.08995} & {0.07594} & {0.18552} & {0.12951} \\
\cmidrule(lr){2-7} \cmidrule(lr){8-13}
  \textsc{UserPop} & {0.04697} & {0.06147} & {0.06549} & 0.03814 & 0.09697 & 0.07890 & \best{0.17776} & \best{0.21648} & \best{0.22895} & \best{0.13901} & \best{0.32613} & \best{0.26999} \\
  \textsc{ItemPop} & \secondbest{0.05954} & \secondbest{0.07433} & \secondbest{0.08456} & \secondbest{0.04292} & \secondbest{0.10473} & \secondbest{0.09410} & {0.06008} & {0.07519} & {0.08505} & {0.04680} & {0.11625} & {0.09928} \\
  \textsc{GlobalPop} & 0.04953 & 0.06377 & 0.06596 & {0.03305} & {0.08487} & {0.07172} & 0.10596 & 0.14132 & 0.13923 & 0.09357 & 0.24662 & 0.18156 \\
\cmidrule(lr){2-7} \cmidrule(lr){8-13}
  \textsc{BPER+} & 0.05084 & 0.06550 & 0.07313 & 0.03454 & 0.08893 & 0.07984 & 0.10716 & 0.14349 & 0.14300 & \secondbest{0.09417} & \secondbest{0.24909} & 0.18545 \\
  \;\;\; \textit{Improv.} & \red{-0.00870} & \red{-0.00883} & \red{-0.01142} & \red{-0.00839} & \red{-0.01580} & \red{-0.01426} & \red{-0.07060} & \red{-0.07299} & \red{-0.08595} & \red{-0.04485} & \red{-0.07705} & \red{-0.08454} \\
\cmidrule(lr){2-7} \cmidrule(lr){8-13}
  \textsc{ExpGCN} & \best{0.06309}$^{**}$ & \best{0.08187}$^{***}$ & \best{0.08921}$^{**}$ & \best{0.04509}$^{**}$ & \best{0.11686}$^{***}$ & \best{0.10054}$^{***}$ & \secondbest{0.11154}$^{***}$ & \secondbest{0.14791}$^{***}$ & \secondbest{0.14873}$^{***}$ & 0.09279$^{***}$ & 0.24393$^{***}$ & \secondbest{0.18618}$^{***}$ \\
  \;\;\; \textit{Improv.} & \green{0.00355} & \green{0.00754} & \green{0.00466} & \green{0.00216} & \green{0.01213} & \green{0.00643} & \red{-0.06622} & \red{-0.06857} & \red{-0.08022} & \red{-0.04622} & \red{-0.08220} & \red{-0.08381} \\
\midrule
\multicolumn{13}{c}{\textbf{Clothes}} \\
\midrule
  \textsc{Random} & {0.00002} & {0.00003} & {0.00004} & {0.00002} & {0.00005} & {0.00005} & {0.09012} & {0.11141} & {0.10683} & {0.09123} & {0.22660} & {0.15632} \\
\cmidrule(lr){2-7} \cmidrule(lr){8-13}
  \textsc{UserPop} & {0.04908} & {0.06830} & {0.07484} & {0.03416} & {0.09436} & {0.08464} & \best{0.15136} & \secondbest{0.19828} & \best{0.20757} & 0.12156 & 0.31049 & 0.25325 \\
  \textsc{ItemPop} & 0.07143 & 0.09685 & 0.11102 & 0.04646 & 0.12454 & 0.12000 & {0.07290} & {0.09880} & {0.11244} & {0.05376} & {0.14419} & {0.12947} \\
  \textsc{GlobalPop} & \secondbest{0.08777} & \secondbest{0.12130} & \secondbest{0.13154} & \secondbest{0.05905} & \secondbest{0.15790} & \secondbest{0.14494} & 0.14545 & 0.19628 & 0.20035 & \best{0.12225} & \best{0.32373} & \secondbest{0.25389} \\
\cmidrule(lr){2-7} \cmidrule(lr){8-13}
  \textsc{BPER+} & 0.07143 & 0.09877 & 0.08350 & 0.05246 & 0.14202 & 0.10118 & 0.14423 & 0.19519 & 0.19282 & \secondbest{0.12173} & \secondbest{0.32321} & 0.24685 \\
  \;\;\; \textit{Improv.} & \red{-0.01634} & \red{-0.02252} & \red{-0.04805} & \red{-0.00659} & \red{-0.01589} & \red{-0.04376} & \red{-0.00713} & \red{-0.00309} & \red{-0.01476} & \red{-0.00052} & \red{-0.00052} & \red{-0.00705} \\
\cmidrule(lr){2-7} \cmidrule(lr){8-13}
  \textsc{ExpGCN} & \best{0.09647}$^{***}$ & \best{0.13152}$^{***}$ & \best{0.14499}$^{***}$ & \best{0.06469}$^{***}$ & \best{0.17267}$^{***}$ & \best{0.15976}$^{***}$ & \secondbest{0.14782}$^{**}$ & \best{0.19912} & \secondbest{0.20518} & 0.12164 & 0.32241 & \best{0.25646} \\
  \;\;\; \textit{Improv.} & \green{0.00870} & \green{0.01022} & \green{0.01345} & \green{0.00564} & \green{0.01477} & \green{0.01481} & \red{-0.00354} & \green{0.00084} & \red{-0.00239} & \red{-0.00061} & \red{-0.00133} & \green{0.00256} \\
\midrule
\multicolumn{13}{c}{\textbf{Beauty}} \\
\midrule
  \textsc{Random} & {0.00197} & {0.00264} & {0.00413} & {0.00098} & {0.00264} & {0.00392} & {0.06580} & {0.08315} & {0.07941} & {0.06682} & {0.16872} & {0.11619} \\
\cmidrule(lr){2-7} \cmidrule(lr){8-13}
  \textsc{UserPop} & {0.02628} & {0.03503} & {0.03842} & {0.01885} & {0.05036} & {0.04358} & \best{0.10783} & \secondbest{0.14061} & \best{0.14482} & \secondbest{0.09054} & 0.23200 & \best{0.18171} \\
  \textsc{ItemPop} & \secondbest{0.05862} & \secondbest{0.07545} & \secondbest{0.08332} & \secondbest{0.04099} & \secondbest{0.10350} & \secondbest{0.09264} & {0.05895} & {0.07577} & {0.08357} & {0.04468} & {0.11374} & {0.09731} \\
  \textsc{GlobalPop} & 0.03348 & 0.04365 & 0.04527 & 0.02596 & 0.06886 & 0.05478 & 0.09706 & 0.12948 & 0.12475 & 0.08687 & 0.22809 & 0.16637 \\
\cmidrule(lr){2-7} \cmidrule(lr){8-13}
  \textsc{BPER+} & 0.04362 & 0.05827 & 0.05971 & 0.03179 & 0.08327 & 0.06910 & 0.09873 & 0.13233 & 0.12842 & 0.08840 & \secondbest{0.23327} & 0.17112 \\
  \;\;\; \textit{Improv.} & \red{-0.01500} & \red{-0.01718} & \red{-0.02361} & \red{-0.00920} & \red{-0.02022} & \red{-0.02353} & \red{-0.00911} & \red{-0.00828} & \red{-0.01640} & \red{-0.00214} & \green{0.00126} & \red{-0.01059} \\
\cmidrule(lr){2-7} \cmidrule(lr){8-13}
  \textsc{ExpGCN} & \best{0.06200}$^{***}$ & \best{0.08091}$^{***}$ & \best{0.08632} & \best{0.04564}$^{***}$ & \best{0.11808}$^{***}$ & \best{0.10001}$^{***}$ & \secondbest{0.10741} & \best{0.14272} & \secondbest{0.14273} & \best{0.09056} & \best{0.23750}$^{*}$ & \secondbest{0.18159} \\
  \;\;\; \textit{Improv.} & \green{0.00338} & \green{0.00546} & \green{0.00300} & \green{0.00464} & \green{0.01458} & \green{0.00737} & \red{-0.00042} & \green{0.00211} & \red{-0.00209} & \green{0.00002} & \green{0.00550} & \red{-0.00011} \\
\midrule
\multicolumn{13}{c}{\textbf{Sports}} \\
\midrule
  \textsc{Random} & {0.00001} & {0.00000} & {0.00001} & {0.00000} & {0.00000} & {0.00001} & {0.06865} & {0.07906} & {0.07999} & {0.06899} & {0.15810} & {0.11322} \\
\cmidrule(lr){2-7} \cmidrule(lr){8-13}
  \textsc{UserPop} & {0.01966} & {0.02470} & {0.02683} & {0.01697} & {0.04250} & {0.03434} & \best{0.10052} & \secondbest{0.11888} & \best{0.12858} & \best{0.08776} & \best{0.20677} & \best{0.16326} \\
  \textsc{ItemPop} & \secondbest{0.05103} & \secondbest{0.06277} & \secondbest{0.07071} & \secondbest{0.03626} & \secondbest{0.08859} & \secondbest{0.07921} & {0.05144} & {0.06330} & {0.07112} & {0.03893} & {0.09542} & {0.08252} \\
  \textsc{GlobalPop} & 0.04016 & 0.05036 & 0.05322 & 0.03075 & 0.07589 & 0.06309 & 0.09472 & 0.11662 & 0.11694 & \secondbest{0.08481} & \secondbest{0.20418} & 0.15370 \\
\cmidrule(lr){2-7} \cmidrule(lr){8-13}
  \textsc{BPER+} & 0.01340 & 0.01781 & 0.01502 & 0.01529 & 0.03939 & 0.02486 & 0.08786 & 0.10870 & 0.10429 & 0.08282 & 0.20070 & 0.14371 \\
  \;\;\; \textit{Improv.} & \red{-0.03763} & \red{-0.04495} & \red{-0.05569} & \red{-0.02098} & \red{-0.04920} & \red{-0.05434} & \red{-0.01266} & \red{-0.01017} & \red{-0.02428} & \red{-0.00494} & \red{-0.00606} & \red{-0.01955} \\
\cmidrule(lr){2-7} \cmidrule(lr){8-13}
  \textsc{ExpGCN} & \best{0.05547}$^{***}$ & \best{0.06814}$^{***}$ & \best{0.07593}$^{***}$ & \best{0.04055}$^{***}$ & \best{0.09852}$^{***}$ & \best{0.08639}$^{***}$ & \secondbest{0.09930} & \best{0.12128} & \secondbest{0.12775} & 0.08373$^{***}$ & 0.20148 & \secondbest{0.15978}$^{*}$ \\
  \;\;\; \textit{Improv.} & \green{0.00444} & \green{0.00538} & \green{0.00522} & \green{0.00429} & \green{0.00994} & \green{0.00718} & \red{-0.00122} & \green{0.00240} & \red{-0.00082} & \red{-0.00403} & \red{-0.00529} & \red{-0.00348} \\
\bottomrule
\end{tabular}
}
\label{tab:statement-ranking-results}
\vspace{-2mm}
\end{center}
\end{table*}

\subsection{\textsc{StaR} Benchmark Datasets}
We construct the \textbf{Sta}tement \textbf{R}anking benchmark, abbreviated \textsc{StaR}, by applying our statement extraction and paraphrase clustering pipeline to four different categories of the Amazon Reviews 2014 dataset~\cite{ni2019justifying}~\footnote{\url{https://jmcauley.ucsd.edu/data/amazon/links.html}}.
We consider the following categories: \textbf{Toys and Games} (\textit{Toys}), \textbf{Clothing, Shoes and Jewelry} (\textit{Clothes}), \textbf{Beauty} (\textit{Beauty}), and \textbf{Sports and Outdoors} (\textit{Sports}).
These domains are widely used in the recommendation literature and provide diverse product contexts for evaluation.
For each dataset, we extract statements from each review and apply clustering at the dataset level.
We retain interactions with at least one statement, resulting in datasets with between 115K and 294K interactions and between 718K and 1.3M user--item--statement triplets after clustering.
We apply a temporal split per user, using the last interaction for testing, the second-to-last for validation, and all remaining interactions for training. 
Complete dataset statistics, including details on statements before and after clustering, are reported in Table~\ref{tab:dataset_stats}.

\section{Experiments}
We conduct our experiments in two parts: we first evaluate statement-ranking models on \textsc{StaR}, and then assess benchmark quality by measuring the statement properties induced by our extraction and clustering procedures, using both human annotations and automatic evaluation.

\subsection{Statement Ranking Evaluation on \textsc{StaR}}
We introduce popularity-based baselines and evaluate them on \textsc{StaR} alongside two state-of-the-art methods under two complementary settings, global-level and item-level ranking.
We also study the sensitivity of each model to key hyperparameters and compare performance trends across the two ranking paradigms.

\subsubsection{Methods}
We evaluate six statement ranking methods on \textsc{StaR}: the \textsc{Random} baseline, three popularity methods (\textsc{UserPop}, \textsc{ItemPop}, \textsc{GlobalPop}), and two state-of-the-art approaches for explanatory passage ranking (\textsc{BPER+}~\cite{li2023relationship}, \textsc{ExpGCN}~\cite{wei2023expgcn}). 


\subsubsection*{\textbf{Popularity-based Methods}}
We introduce three simple baselines based on statement frequency in the training data:
\begin{itemize}[leftmargin=*]
\item \textit{User-based Popularity} (\textsc{UserPop}): ranks statements by their frequency in the user's interactions:
\begin{align}
    \hat{r}_{u,i,s} = \sum_{i' \in \mathcal{I}_u} \delta(s \in \mathcal{S}_{ui'})
\end{align}
where $\mathcal{I}_u$ is the set of items with which user $u$ has interacted.
\item \textit{Item-based Popularity} (\textsc{ItemPop}): ranks statements by their frequency in the item's interactions:
\begin{align}
    \hat{r}_{u,i,s} = \sum_{u' \in \mathcal{U}_i} \delta(s \in \mathcal{S}_{u'i})
\end{align}
where $\mathcal{U}_i$ is the set of users who have interacted with item $i$.
\item \textit{Global Popularity} (\textsc{GlobalPop}): ranks statements by their frequency across all interactions:
\begin{align}
    \hat{r}_{u,i,s} = \sum_{i' \in \mathcal{I}} \sum_{u' \in \mathcal{U}_{i'}} \delta(s \in \mathcal{S}_{u'i'})
\end{align}
\end{itemize}

\subsubsection*{\textbf{State-of-the-art Explanation Ranking Methods}}
\textsc{BPER+} models the ternary user--item--statement relation by decomposing it into user- and item--statement interactions, enriching statement representations with \textsc{BERT}~\cite{devlin2019bert}, and combining user- and item-side relevance via a weight $\mu$. \textsc{ExpGCN} learns explanation-aware representations from the user--item--statement graph by applying graph convolutions on homogeneous subgraphs, and computes relevance by summing the user- and item--statement scores.

\subsubsection{Experimental Setup}
We implement the random and popularity baselines and reuse the official implementations for \textsc{BPER+}~\footnote{\url{https://github.com/lileipisces/BPER}} and \textsc{ExpGCN}~\footnote{\url{https://github.com/Joinn99/ExpGCN}}.
For \textsc{BPER+}, we use \texttt{Bert-base-uncased}~\footnote{\url{https://huggingface.co/google-bert/bert-base-uncased}} for statement embeddings and set $d = 20$ (model dimension), $\gamma = 10^{-5}$ (learning rate), and $\mu = 0.7$ (user--item score weighting) after hyperparameter tuning across all datasets.
For \textsc{ExpGCN}, we set $d = 128$, $\gamma = 10^{-3}$, and $L = 4$ (number of layers) for user--statement and item--statement and $2$ for the user--item subgraphs.
We analyze the impact of key hyperparameters ($\mu$ for \textsc{BPER+} and $L$ for \textsc{ExpGCN}) on model performance.
Other hyperparameters follow the original implementations.
Training is capped at 500 epochs for each model, and we report the average performance over 3 runs, selecting the best model on the validation set using NDCG@10.

\subsubsection{Global-Level Ranking Results}
Table~\ref{tab:statement-ranking-results} (Global-level) presents global-level ranking performance.
\textsc{ExpGCN} consistently achieves the best results across all datasets and metrics, improving NDCG@10 over the strongest baseline by $+0.00643$ on \textit{Toys} and up to $+0.01481$ on \textit{Clothes}.
Popularity baselines remain highly competitive: \textsc{ItemPop} is generally the strongest (except on \textit{Clothes}, where \textsc{GlobalPop} leads), and \textsc{GlobalPop} consistently outperforms \textsc{UserPop}.
In contrast, \textsc{BPER+} underperforms the strongest baseline on every dataset, with the largest degradation on \textit{Sports} ($-0.05434$ on NDCG@10). This suggests that, in the global setting, item- and global-level signals dominate, and an imperfect balance between user- and item-side relevance can substantially degrade ranking quality.
Finally, \textsc{Random} is near-zero on \textit{Clothes} and \textit{Sports} but noticeably higher on \textit{Toys} and \textit{Beauty}, reflecting dataset-dependent difficulty induced by differences in the effective candidate space.
Overall, global-level ranking mainly emphasizes retrieval from a large statement universe. Graph-based modeling of the user--item--statement structure, combined with strong frequency priors, yields consistent gains.

\subsubsection{Item-Level Ranking Results}
Table~\ref{tab:statement-ranking-results} (Item-level) shows a different trend from the global-level setting.
\textsc{UserPop} becomes a particularly strong baseline in \textit{Sports} and, even more clearly, in \textit{Toys}, where \textsc{ExpGCN} exhibits significant negative improvements over \textsc{UserPop} in all metrics ($-0.08381$ in NDCG@10). On \textit{Sports}, \textsc{ExpGCN} remains closer to \textsc{UserPop} and even yields small positive gains on some metrics ($+0.00240$ on R@5).
In contrast, \textsc{ItemPop} collapses in the item-level setting and is outperformed by \textsc{Random} across all datasets.
The superiority of \textsc{UserPop} is dataset-dependent: on \textit{Clothes}, the best results are split between \textsc{GlobalPop} and \textsc{ExpGCN} depending on the metric.
\textsc{BPER+} is generally outperformed by \textsc{UserPop}. 
Overall, item-level ranking exposes a personalization gap: strong global-level models do not systematically translate item filtering into better user-specific re-ranking, while user-history frequency provides a consistently strong signal.

\subsubsection{Hyperparameter Sensitivity Analysis}
\begin{figure}[h]
    \centering
    \includegraphics[width=\linewidth]{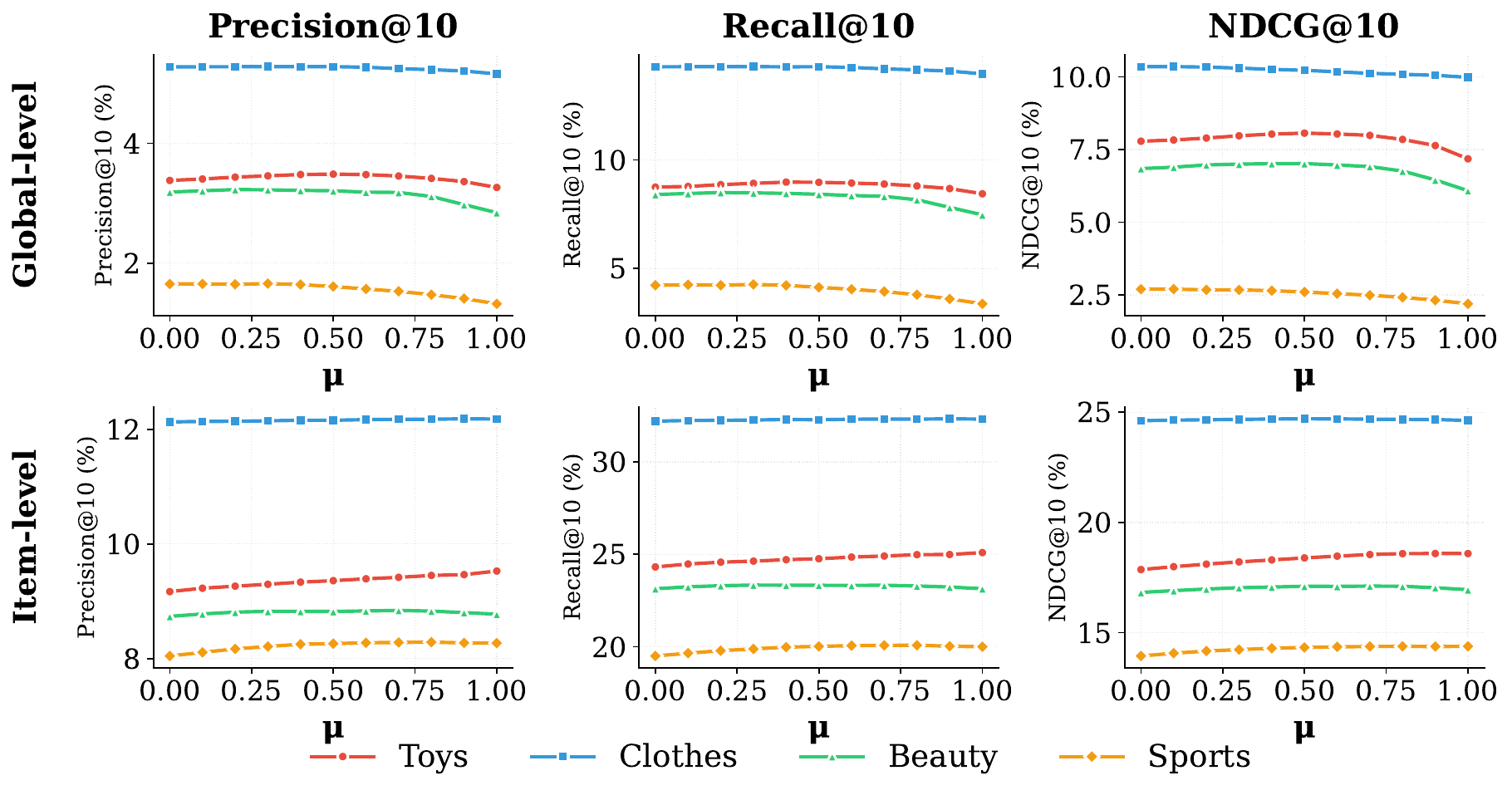}
    \vspace{-6mm}
    \caption{Impact of the $\mu$ parameter on BPER+ performance across datasets. 
    }
    \vspace{-4mm}
    \label{fig:bperp_mu_impact}
    \Description{Impact of the $\mu$ parameter on BPER+ performance across datasets. The top row shows global-level metrics while the bottom row presents item-level metrics. Each column corresponds to a different ranking metric: Precision@10, Recall@10, and NDCG@10.}
\end{figure}

\begin{figure}[h]
    \centering
    \includegraphics[width=\linewidth]{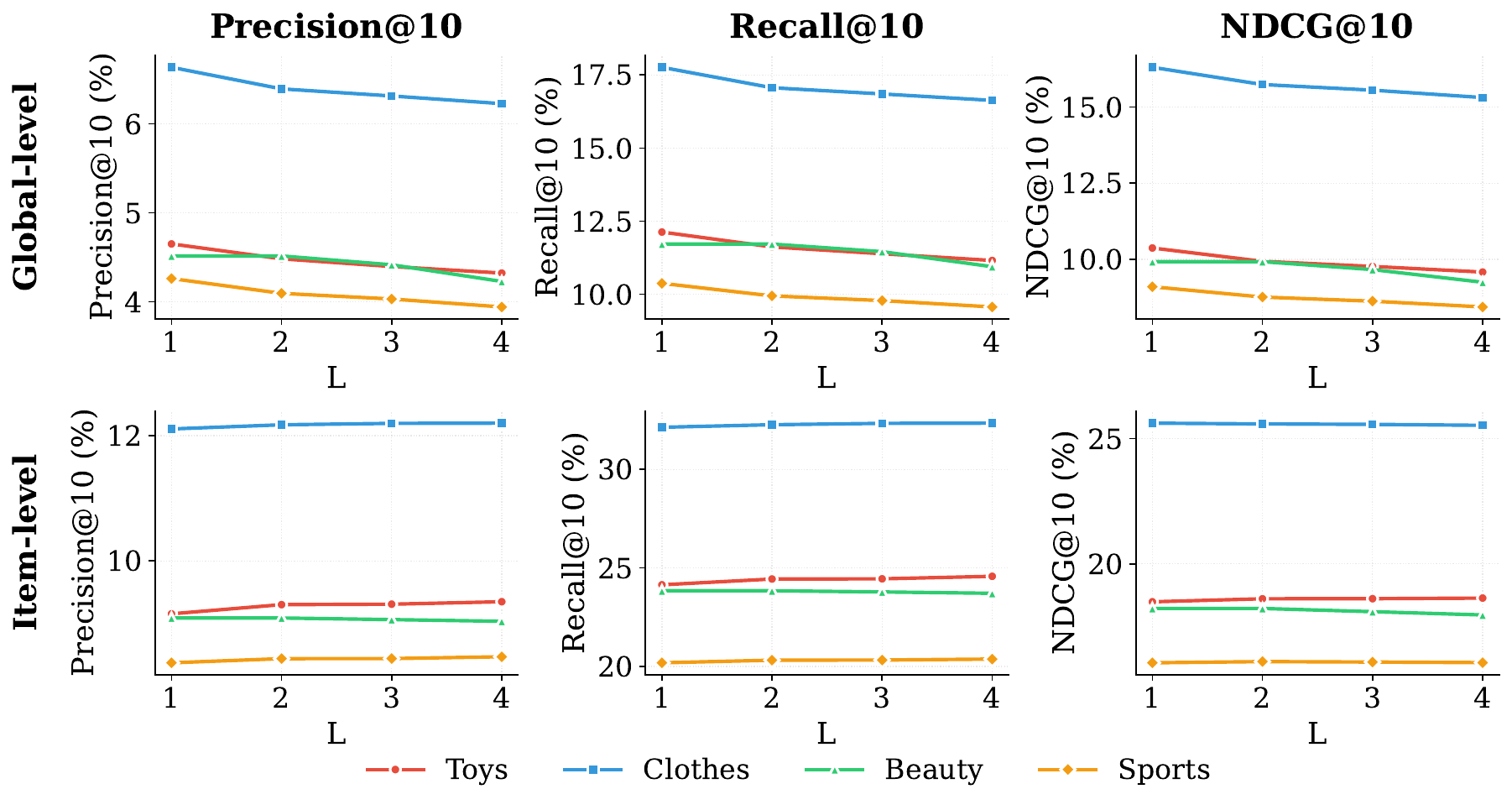}
    \vspace{-6mm}
    \caption{Impact of the number of layers $L$ on ExpGCN performance across datasets. 
    }
    \label{fig:expgcn_l_impact}
    \Description{Impact of the number of layers $L$ on ExpGCN performance across datasets. The top row shows global-level metrics while the bottom row presents item-level metrics. Each column corresponds to a different ranking metric: Precision@10, Recall@10, and NDCG@10.}
\end{figure}

We analyze the impact of key hyperparameters on ranking performance.
Figure~\ref{fig:bperp_mu_impact} reports the effect of $\mu$ in \textsc{BPER+}. In global-level ranking, performance typically peaks at intermediate values, while extreme weights degrade results, with metrics dropping as $\mu$ tends to $1$. In item-level ranking, larger $\mu$ values monotonically improve performance, suggesting stronger benefits from emphasizing the user-side signal, whereas global-level ranking favors a more balanced user--item combination.
Figure~\ref{fig:expgcn_l_impact} shows the sensitivity of \textsc{ExpGCN} to the number of graph convolution layers $L$. In global-level ranking, increasing $L$ consistently hurts performance across datasets, indicating that deeper propagation is detrimental. In item-level ranking, the effect is weaker: performance is largely stable, with small gains on \textit{Toys} and marginal changes elsewhere.

\subsection{\textsc{StaR} Benchmark Quality Evaluation}
We evaluate \textsc{StaR} quality through required statement properties: \textbf{human and LLM-based evaluation} for \textit{explanatoriness} and \textit{atomicity}, and \textbf{unsupervised clustering metrics} for \textit{uniqueness}.
We also provide qualitative comparisons with \textsc{Extra} and illustrate example clusters.

\subsubsection{Statement Extraction Analysis}
We assess extraction quality through \emph{explanatoriness} and \emph{atomicity}.
For each dataset, we randomly sample 75 interactions for human evaluation and 10{,}000 for automatic evaluation with \texttt{Llama-3.1-8B-Instruct}~\cite{dubey2024llama}~\footnote{\url{https://huggingface.co/meta-llama/Llama-3.1-8B-Instruct}}.
We compute, for each interaction, the proportion of extracted statements judged (i) explanatory with respect to the source review, and (ii) atomic, then average across interactions.
We report results in Table~\ref{tab:statement_quality}, both before and after the verification step of our extraction pipeline.
With verification enabled, human evaluation indicates that $92.16\%$--$96.09\%$ of considered statements are explanatory, and $80.92\%$--$93.87\%$ are atomic.
The LLM-as-a-judge evaluation yields comparable estimates, with $93.41\%$--$97.20\%$ explanatory statements and $91.36\%$--$95.50\%$ atomic statements.
Overall, the verification step consistently improves statement quality, with gains of up to $10\%$ compared to extraction without verification, for both explanatoriness and atomicity, supporting the value of this additional filtering stage.

\begin{table}[t]
\centering
\caption{Statement quality evaluation on \textsc{StaR} benchmark. We report the proportion (\%) of statements satisfying the \emph{Explainatoriness} and \emph{Atomicity} properties, with (\cmark) and without (\xmark) the verification step, and report improvement.}
\vspace{-2mm}
\label{tab:statement_quality}
\resizebox{0.9\linewidth}{!}{
\begin{tabular}{lrrrrrr}
\toprule
& \multicolumn{3}{c}{\textbf{Explainatoriness}} & \multicolumn{3}{c}{\textbf{Atomicity}} \\
\cmidrule(lr){2-4}\cmidrule(lr){5-7}
{Verification} & \xmark & \cmark & \textit{Improv.} & \xmark & \cmark & \textit{Improv.} \\
\midrule
\multicolumn{7}{c}{\textbf{Human Evaluation}} \\
\midrule
\textbf{Toys}    & 79.71 & 93.93 & \green{+14.22} & 71.53 & 83.75 & \green{+12.22} \\
\textbf{Clothes} & 83.46 & 95.65 & \green{+12.19} & 86.35 & 93.87 & \green{+7.52} \\
\textbf{Beauty}  & 84.09 & 92.16 & \green{+8.07} & 66.75 & 80.92 & \green{+14.17} \\ 
\textbf{Sports}  & 85.12 & 96.09 & \green{+10.97} & 79.97 & 88.48 & \green{+8.51} \\
\midrule
\multicolumn{7}{c}{\textbf{LLM-based Evaluation}} \\
\midrule
\textbf{Toys}    & 82.70 & 93.41 & \green{+10.71} & 80.37 & 91.36 & \green{10.99} \\
\textbf{Clothes} & 86.66 & 94.62 & \green{+7.96} & 84.16 & 92.86 & \green{+8.7} \\
\textbf{Beauty}  & 82.66 & 93.51 & \green{+10.85} & 80.30 & 91.40 & \green{+11.1} \\
\textbf{Sports}  & 92.63 & 97.20 & \green{+4.57} & 90.21 & 95.50 & \green{+5.29} \\
\bottomrule
\end{tabular}
}
\vspace{-2mm}
\end{table}

\begin{table}[t]
\centering
\caption{Qualitative comparison of statement extraction on \textit{Toys} between \textsc{Extra} and \textsc{StaR}.}
\vspace{-2mm}
\small
\resizebox{\linewidth}{!}{
\begin{tabular}{p{0.45\linewidth} p{0.12\linewidth} p{0.43\linewidth}}
\toprule
\textbf{Review} & \textbf{\textsc{Extra}} & \textbf{\textsc{StaR} (statement, sentiment)}\\
\midrule
My grand daughter has had so much fun with this set. She plays for hours and enjoys every minute. Good Job! Great quality toy. &
\textit{good job} &
the product is enjoyable to use (\textsc{pos}); the product is of great quality (\textsc{pos}); the product is a toy (\textsc{neu})\\
\midrule
Bought for my 5 year old nephew. Great toy!! He loves it! Stickers were great and easy to open the mystery machine. &
\textit{great toy} &
the stickers look great (\textsc{pos}); the product is easy to open (\textsc{pos}); the product is a toy (\textsc{neu}) \\
\bottomrule
\end{tabular}
}
\label{tab:extraction_qualitative_examples}
\vspace{-2mm}
\end{table}

In Table~\ref{tab:extraction_qualitative_examples}, we qualitatively compare statements extracted by \textsc{Extra} and \textsc{StaR} on a few representative reviews.
The examples highlight several limitations of the heuristic selection strategy used in \textsc{Extra}: many review sentences are not explanatory, some require rewriting or abstraction to surface the underlying rationale, and filtering sentences with personal pronouns can discard genuinely explanatory content.
Moreover, even highly recurrent snippets (e.g., \emph{good job}) may carry little explanatory information despite being frequent.
In contrast, \textsc{StaR} leverages LLM language understanding to extract, from noisy reviews, their explanatory substance in the form of atomic statements paired with sentiment labels.

\subsubsection{Statement Clustering Analysis}
We evaluate clustering quality in \textsc{StaR} to quantify statement \emph{uniqueness}.
Since exhaustive pairwise assessment is infeasible, we rely on unsupervised embedding-based metrics, within-cluster dispersion (SSE) and between-cluster separation (SSB)~\cite{palacio2019evaluation}, using $d(\cdot)=1-\text{\textsc{Cosine}}(\cdot)$.
We compare the $n$-gram LSH clustering used in \textsc{Extra}~\cite{anand2011mining} with our semantic pipeline and ablations (Table~\ref{tab:clustering_quality}), computing all metrics with \texttt{KaLM-Embedding-Gemma3-12B-2511}~\cite{zhao2025kalm}~\footnote{\url{https://huggingface.co/tencent/KaLM-Embedding-Gemma3-12B-2511}}.
Our complete pipeline achieves $40.43\%$--$54.80\%$ reduction and the best SSB across datasets, indicating better semantic separation, whereas LSH yields near-zero SSE but an almost trivial clustering with at most $0.22\%$ reduction, failing to merge paraphrases beyond lexical overlap.
Overall, the complete pipeline produces the most cohesive and well-separated clusters, and we show representative examples in Table~\ref{tab:cluster-examples}.
\begin{table}[t]
\centering
\caption{Clustering quality evaluation. We report reduction statistics and unsupervised metrics (SSE, SSB).}
\vspace{-2mm}
\label{tab:clustering_quality}
\resizebox{0.95\linewidth}{!}{
\begin{tabular}{@{}llrrrr@{}}
\toprule
\textbf{Dataset} & \textbf{Method} & {\#Clust.} & {Red.\%} & {SSE $\downarrow$} & {SSB $\uparrow$} \\
\midrule
\textbf{Toys}   & N-gram + LSH              & 472\,019 & \red{0.17} & \textbf{0.0000} & \underline{0.0785} \\
       & ANN                       & 225\,270 & 52.35      & 0.0516          & 0.0751 \\
       & ANN + Filter              & 229\,262 & 51.51      & 0.0492          & 0.0754 \\
       & ANN + Filter + Refine     & 281\,664 & 40.43      & \underline{0.0033} & \textbf{0.0818} \\
\midrule
\textbf{Clothes}& N-gram + LSH              & 569\,103 & \red{0.22} & \textbf{0.0000} & \underline{0.0713} \\
       & ANN                       & 173\,882 & 69.51      & 0.1226          & 0.0672 \\
       & ANN + Filter              & 178\,722 & 68.66      & 0.1167          & 0.0675 \\
       & ANN + Filter + Refine     & 260\,477 & 54.33      & \underline{0.0068} & \textbf{0.0748} \\
\midrule
\textbf{Beauty} & N-gram + LSH              & 506\,869 & \red{0.16} & \textbf{0.0000} & 0.0725 \\
       & ANN                       & 154\,755 & 69.51      & 0.1249          & 0.0734 \\
       & ANN + Filter              & 158\,744 & 68.73      & 0.1196          & \underline{0.0735} \\
       & ANN + Filter + Refine     & 229\,448 & 54.80      & \underline{0.0065} & \textbf{0.0760} \\
\midrule
\textbf{Sports} & N-gram + LSH              & 942\,442 & \red{0.14} & \textbf{0.0000} & \underline{0.0810} \\
       & ANN                       & 430\,541 & 54.31      & 0.0676          & 0.0766 \\
       & ANN + Filter              & 440\,441 & 53.26      & 0.0640          & 0.0770 \\
       & ANN + Filter + Refine     & 556\,209 & 40.98      & \underline{0.0039} & \textbf{0.0836} \\
\bottomrule
\end{tabular}
}
\vspace{-2mm}
\end{table}

\begin{table}[t]
\centering
\small
\setlength{\tabcolsep}{4pt}
\caption{Qualitative examples of \textsc{StaR} statement clustering on \textit{Toys}. Each block shows a representative statement and a few members from the same cluster.}
\vspace{-2mm}
\resizebox{\linewidth}{!}{
\begin{tabular}{p{0.98\linewidth}}
\toprule
\textbf{Cluster 1: Preschoolers} \\
\textbf{Repr.} the product is good for preschoolers \\
\textbf{Members} the product is perfect for preschoolers; the product is suitable for preschool; the product is good for preschool age \\
\midrule
\textbf{Cluster 2: Breaks after use} \\
\textbf{Repr.} the product breaks after some use \\
\textbf{Members} the product breaks after a few uses; the product broke as soon as it is used; the product fell apart after a few months of use \\
\midrule
\textbf{Cluster 3: Encourages imagination} \\
\textbf{Repr.} the product encourages a child's imagination \\
\textbf{Members} the product encourages children's imagination; the product encourages imaginative thinking; the product encourages a child to use their imagination \\
\bottomrule
\end{tabular}
}
\label{tab:cluster-examples}
\vspace{-2mm}
\end{table}

\subsubsection{Residual noise}
Despite the overall quality of the benchmark, we identified a few failure modes.
These arise from the combination of LLM-based extraction and semantic clustering: extraction may hallucinate facts, contradict the source review, or over-abstract its content, while clustering may merge statements that are close but not strictly equivalent.
As a result, \textsc{StaR} may contain some hallucinated, inconsistent, or approximate rationales.
Table~\ref{tab:star_failure_case} reports one such example.
These cases do not invalidate the benchmark, but clarify the expected residual noise in a large-scale LLM-derived resource.
\begin{table}[t]

\centering
\caption{\textsc{StaR} failure case: extracted statements hallucinate or contradict facts from the review.}
\vspace{-2mm}
\small
\resizebox{\linewidth}{!}{
\begin{tabular}{p{0.49\linewidth} p{0.49\linewidth}}
\toprule
\textbf{Review} & \textbf{Statement}\\
\midrule
Perfect size puzzle for 3 yr old. 
My daughter really likes these Melissa and Doug 12 piece puzzles and can pretty much do it 90\% on her own now, sometimes completely on her own. 
I think any bigger would frustrate her and any smaller would be too easy. 
They come in a variety of pictures and she likes all of them. 
&
\textit{the product is suitable for children aged 3--5}; 

\textit{the product is too difficult for a 3 year old}; 

\textit{the product is too easy for a 5 year old};

\textit{the product is a 1,000 piece jigsaw puzzle}; 

the product has lots of different pictures; 
\\
\bottomrule
\end{tabular}
}
\label{tab:star_failure_case}
\vspace{-2mm}
\end{table}

\section{Discussion}
\subsubsection*{Statement Ranking against Explanation Generation}
Replacing free-form generation with statement-level ranking addresses three core issues.
First, generation often yields hallucinated or generic rationales, whereas ranking restricts explanations to review-grounded statements, eliminating hallucination by construction.
Second, evaluating generated text is hard to standardize and reproduce across lexical, semantic, learned, and LLM-as-a-judge metrics, whereas ranking supports stable evaluation with well-established information retrieval metrics.
Third, paragraph explanations conflate multiple factors without exposing their contribution, while ranking decomposes explanations into atomic units and uses relevance scores to model factor importance, enabling fine-grained analysis of what justifies a recommendation and to what extent.

\subsubsection*{\textsc{StaR} for Statement Ranking}
\textsc{StaR} makes statement-level ranking practically meaningful by ensuring the candidate statement set is \emph{explanatory}, \emph{atomic}, and \emph{unique}. We enforce explanatoriness and atomicity with an LLM extraction plus verification pipeline, which substantially improves both properties, and we enforce uniqueness with scalable semantic clustering, preserving cohesive, well-separated clusters. Together, these steps yield a benchmark where standard ranking metrics reliably reflect explanation quality and measure whether models identify the most relevant justifications for each user--item interaction.

\subsubsection*{Global and Item-level Ranking Comparison}
Our results reveal a strong mismatch between the two evaluation settings.
Under global-level ranking, \textsc{ExpGCN} performs best, yet simple item- and global-frequency baselines (\textsc{ItemPop}, \textsc{GlobalPop}) remain competitive while the user-history signal (\textsc{UserPop}) largely collapses, suggesting a large-scale retrieval regime where item priors dominate.
In contrast, item-level ranking flips the picture: \textsc{UserPop} becomes the strongest baseline on average and can match or surpass \textsc{ExpGCN}, whereas \textsc{ItemPop} drops sharply, indicating a primarily personalization driven regime where item relevance is largely satisfied by construction and the challenge is to prioritize user-specific factors.
Hyperparameter trends support this interpretation: for \textsc{BPER+}, global-level performance peaks at intermediate user--item mixing, while item-level performance improves as the user weight increases.
Overall, these differences expose a clear personalization gap, where gains in global-level ranking do not necessarily translate to better item-level explanation ranking and current models struggle to leverage interaction-specific signals beyond coarse user priors.

\subsubsection*{Limitations and Future Work}
While \textsc{StaR} enforces key statement properties, it relies on an automatic extraction and clustering pipeline, so residual errors and imperfect paraphrase consolidation may introduce noise in supervision and evaluation, and the benchmark may inherit biases from review data.
Beyond binary relevance, extending \textsc{StaR} with graded relevance (e.g., via review prominence or user annotations) could support more nuanced evaluation, and incorporating diversity and coverage objectives could better reflect the multi-faceted nature of explanations.
Finally, the strong item-level results of simple user-history signals highlight a personalization gap that calls for models explicitly learning fine-grained user preferences over explanatory factors.

\section{Conclusion}
In this paper, we advocated a shift in objective for text-based explainable recommendation: \emph{rank, don't generate}. 
By casting explanation as \emph{statement-level ranking} over review-grounded candidates, we mitigate hallucination by construction and enable standardized, reproducible evaluation with established ranking metrics.
To make this evaluation meaningful, we ensured that candidate statements satisfy three key properties (\emph{explanatoriness}: item facts affecting user experience, \emph{atomicity}: one opinion about one aspect, and \emph{uniqueness}: no redundant paraphrases) through an LLM-based two-stage extraction and verification pipeline, complemented by scalable semantic clustering to consolidate paraphrases beyond lexical overlap.
Building on this pipeline, we introduced \textsc{StaR}, a benchmark spanning four Amazon Reviews 2014 categories, and evaluated both global-level ranking (all dataset statements) and item-level ranking (target item statements).
Our results reveal a sharp mismatch between these settings: state-of-the-art models are competitive for global-level retrieval alongside strong frequency priors, yet in the item-level setting, a simple user-history popularity baseline can match or surpass them, exposing a persistent personalization gap for faithful explanation ranking.

\section*{Acknowledgements}
This work was performed using HPC resources from GENCI–IDRIS (Grants 2025-AD011016963 and 2025-AD010616958).

\bibliographystyle{ACM-Reference-Format}
\bibliography{output}

\end{document}